\documentclass[12pt]{article} 
\usepackage[hyperfootnotes=true]{hyperref}
\usepackage{epsfig}
\usepackage{amsmath}
\usepackage{amssymb}
\usepackage{graphicx}
\usepackage{xcolor}
\usepackage{cite} 
\newlength{\abstractwidth}
\usepackage{ wasysym } 
\usepackage{mciteplus} 
\usepackage{subcaption}
\usepackage{bbold}
\usepackage{slashed}

\usepackage{mathbbol}
\DeclareSymbolFontAlphabet{\amsmathbb}{AMSb}%

\usepackage{amsthm}
\usepackage{hyperref}
\usepackage{color}
\usepackage{mathrsfs}
\usepackage{soul}

\usepackage[outline]{contour}

\DeclareMathAlphabet{\mathpzc}{OT1}{pzc}{m}{it}

\usepackage{calligra}
\DeclareMathAlphabet{\mathcalligra}{T1}{calligra}{m}{n}
\DeclareFontShape{T1}{calligra}{m}{n}{<->s*[2.2]callig15}{}

\def\be {\begin{equation}}
\def\ee {\end{equation}}
\def\bea {\begin{eqnarray}}
\def\eea {\end{eqnarray}}
\def\bc {\begin{center}}
\def\ec {\end{center}}
\def\bfg {\begin{figure}}
\def\efg {\end{figure}}
\def\bi {\begin{itemize}}
\def\ei {\end{itemize}}
\def\nn {\nonumber}

\def\la {\label}
\def\le {\left}
\def\ri {\right}

\def\fr {\frac}

\def\a  {\alpha}
\def\b  {\beta}

\def\beq{\begin{equation}}
\def\eeq{\end{equation}}
\def\br{\begin{eqnarray}}
\def\er{\end{eqnarray}}
\newcommand{\eel}[1] {\label{#1}\end{equation}}

\begin{document}

\begin{titlepage}

\begin{center}
\centerline
{\Large \bf { Lorentz and gauge invariance of quantum space  }} 
\bigskip
\end{center}

\begin{center}
{ \bf   Ahmed Farag Ali$^{\triangle 	\nabla}$\footnote{ afaragali@ucmerced.edu; ahmed.ali@fsc.bu.edu.eg}, Barun Majumder$^{\otimes}$\footnote{ barunbasanta@gmail.com} and Prabir Rudra$^{\Box}$\footnote{ prudra.math@gmail.com (corresponding author)}}, 

\bigskip
{\em $^\triangle$ Essex County College, 303 University Ave, Newark, NJ 07102, United States.}\\
{\em $^\nabla$Dept. of Physics, Benha University, Benha 13518, Egypt} \\
{\em $^\otimes$ {University of Tennessee, Knoxville, TN 37996, USA}}\\
{\em $^{\Box}$Department of Mathematics, Asutosh College, Kolkata-700026, India} 

\bigskip

\end{center}

\begin{abstract}
Motivated by the generalized uncertainty principle, we derive a discrete picture of the space that respects Lorentz symmetry as well as gauge symmetry by setting an equivalency between the linear Generalized Uncertainty Principle (GUP) correction term and electromagnetic interaction term in the Dirac equation. We derived a wave function solution that satisfies this equivalency. This discreteness may explain the crystal and quasicrystal structures observed in nature at different energy scales.
\end{abstract}


  \end{titlepage}

\section{Introduction}
Various approaches of quantum gravity predict an existence of minimum measurable length  \cite {Markov:1967lha, Kadyshevsky:1978em, Kadyshevsky:1977mu, Amati:1988tn, Konishi:1989wk, Garay:1994en, Scardigli:1999jh, Capozziello:1999wx, Ali:2009zq, Zhu:2008cg, Mignemi:2009ji, Das:2010zf, Ali:2011fa, Pedram:2011gw, Isi:2013cxa, Shababi:2017zrt, Mureika:2018gxl}. The minimal length and the Generalized Uncertainty Principle (GUP) models attracted experimental and observational studies on atomic physics \cite{Brau:1999uv, Das:2008kaa}, quantum optics \cite{Pikovski:2011zk}, gravitational bar detectors \cite{Marin:2013pga}, quantum gravitational decoherence \cite{Petruzziello:2020wkd}, condensed matter physics \cite{Iorio:2017vtw}, cold atoms \cite{Gao:2016fmk} and gravitational waves \cite{Bosso:2018ckz}. A review on minimal length theories can be found in \cite{Hossenfelder:2012jw}. One of the GUP models shows a discrete picture of the space \cite{Ali:2009zq, Das:2010zf, Ali:2011fa}. It was proved that to confine a particle in a box in any number of dimensions, the dimensions of the box should be quantized as multiples of the fundamental Planck length. The result has been proved for Schrodinger equation \cite{Ali:2009zq, Ali:2011fa}, Klein-Gordon and Dirac equations \cite{Das:2010zf} which implies a universality of space quantization. However, the existence of the fundamental length scale or GUP may be equivalent to modified dispersion relations \cite{Amelino-Camelia:2005zpp} and possible violation of Lorentz symmetry \cite{Lambiase:2017adh}.

In order to have a relativistic model of minimal length, two authors of this study have shown recently in \cite{Ali:2021oml}  that the discreteness of space/minimal measurable length can be obtained due to Dresselhaus anisotropic spin–orbit interaction. This model \cite{Ali:2021oml} only works for particles whose spin and does not work for spin-less particles described by the Klein-Gordon equation. This sets an advantage for linear GUP that universally affect any physical system through its kinetic energy term \cite{Kempf:1994su, Das:2008kaa, Ali:2011fa, Fadel:2021hnx} and implies discreteness of space that confines either spin-less or spin particles.  Since GUP and Dresselhaus anisotropic spin-orbit interaction implies similar discreteness of space with similar solutions, we suggest that the Dresselhaus anisotropic spin-orbit interaction could be a possible manifestation of linear GUP \cite{Ali:2009zq, Ali:2011fa}. This may provide a support to GUP as a robust and universal effect. On the conceptual level, spin-orbit interaction is a relativistic interaction of a particle's spin with its motion inside a potential. This potential can be fundamentally due to one of the four fundamental forces. In this study, we assume that electromagnetic force is a dominating fundamental force that generates spin-orbit interaction. In order to have a discrete picture of space that respect Lorentz symmetry as well as gauge symmetry and is still consistent with the core idea of the universality of GUP, we suggest an equivalency between the GUP effect and electromagnetic potential at the conceptual level. The wave function solution implied by this equivalency describes a physical system that could be spin-less or spin particles that are confined in a discrete space that respects both Lorentz and Gauge symmetry. The question of minimal length and gauge invariance was discussed conceptually in \cite{Chang:2016sae} in which the authors found an interplay between the minimal spatial and magnetic lengths. Our study may be used to explain the crystal and quasicrystal structures that we observe in nature at different energy scales.

In this paper we use the GUP as presented in \cite{Ali:2009zq, Das:2010zf, Ali:2011fa} to study the GUP-corrected Dirac equation in conjunction with Dirac's equation in electromagnetic (EM) field. We argue that the Lorentz invariant EM part of the Hamiltonian is capable of inducing a discrete nature of space in the quantum analog of space subject to its equivalency with GUP \cite{Ali:2009zq, Das:2010zf, Ali:2011fa}. Discrete space in that sense is a possible solution due to quantum electromagnetic interaction.  This may resonate with the emergence of gravity (space) from quantum interactions \cite{dassur, tedprl, Verlinde:2010hp}. The paper is organized as follows: Section 2 is dedicated to the discussion of GUP in the presence of electromagnetic fields, where we have discussed the Dirac equation in one dimension. In section 3 we discuss the Dirac equation in two dimensions. Section 4 deals with the intricacies of the Dirac equation in three dimensions. Finally, the paper ends with a brief discussion and conclusion in section 5.

\section{GUP and EM}

In this section we review the discreteness of space that was implied by linear GUP in \cite{Ali:2009zq}. The linear GUP takes the following form 

\bea
[x_i, p_j] = i \hbar\hspace{-0.5ex} \left[  \delta_{ij}\hspace{-0.5ex}
- \hspace{-0.5ex} a \hspace{-0.5ex}  \le( p \delta_{ij} +
\frac{p_i p_j}{p} \ri)
+ a^2 \hspace{-0.5ex}
\le( p^2 \delta_{ij}  + 3 p_{i} p_{j} \ri) \hspace{-0.5ex} \ri]
\label{comm01}
\eea
where
$p^{2} = \sum\limits_{j=1}^{3}p_{j}p_{j} $, $a = {a_0}/{M_{Pl}c}
= {a_0 \ell_{Pl}}/{\hbar},$
$M_{Pl}=$ Planck mass, $\ell_{Pl}\approx 10^{-35}~m=$ Planck length,
and $M_{Pl} c^2=$ Planck energy $\approx 10^{19}~GeV$.
It was proved in \cite{Ali:2009zq} that the commutation relation in Eq. (\ref{comm01}) would imply minimum measurable length and maximum measurable momentum depending on the value of $a $ parameter.
It was found in \cite{Ali:2009zq, Ali:2011fa} that the momentum and position can be redefined in order to reproduce the commutation relation in Eq. (\ref{comm01}). This modification reads

\bea x_i = x_{0i}~,~~
p_i = p_{0i} \le( 1 - a~p_0 + 2 a^2 p_0^2 \ri)~, \la{mom1}
\eea
%
with $x_{0i}, p_{0j}$
satisfying the standard canonical commutation relations. The momentum $p_{0i}$ is defined as the momentum at low energies with the standard representation in position space. Using (\ref{mom1}), a Hamiltonian of the form
\bea H &=& \fr{p^2}{2m} + V(\vec r)
\eea
can be written as
\bea
H&=&H_0 + H_1 + {\cal O}(a^2) ~, \label{GUPH} \\
%
%
\mbox{where}~H_0 &=& \fr{p_0^2}{2m} + V(\vec r)   \\
\mbox{and}~ H_1 &=& -\fr{a}{m}~p_0^3~. 
\eea
which indicates that the classical or quantum Hamiltonian for any physical system would be corrected by GUP correction. This implies the universality of quantum gravity corrections \cite{Das:2008kaa, Ali:2011fa}. When this correction was studied with Schrodinger equation \cite{Ali:2009zq}, it implied a  third order Schr\"odinger equation that has a new {\it non-perturbative} solution of the form $\psi \sim e^{ix/2a\hbar}$, which when superposed with the regular solutions perturbed by terms ${\cal O}(a)$, implies both quantizations of energy and length at the same time. The discrete picture of space is described by the following equation
\bea
\frac{L}{a\hbar} = \frac{L}{a_0 \ell_{Pl}} = 2p\pi + \theta~,~p \in \mathbb{N}
\la{quant1}
\eea
where $\theta = {\cal O}(1)$. This has been interpreted as the quantization of measurable lengths. When considering the Dirac equation \cite{Das:2010zf}, it implied a discreteness/quantization of the space in terms of the $\alpha$ parameter of linear GUP. The GUP-corrected Dirac equation takes the following form\cite{Das:2010zf}.
\bea
H \psi &=& \le (c\, \vec \a \cdot \vec p + \b mc^2 \ri) \psi (\vec r)  \nn \\
&=&  \le(c\, \vec \a \cdot \vec p_0 -
c\, a (\vec\a \cdot \vec p_0)(\vec\a \cdot \vec p_0) + \b mc^2 \ri) \psi (\vec r)\nn \\
&=& E\psi (\vec r)
\la{ham1}
\eea
Here $\alpha$ and $\beta$ are Dirac matrices. The solutions of the GUP-corrected Dirac equation implied discreteness of length, area, and volume\cite{Das:2010zf}. This indicates that linear GUP introduces a quantum picture of the space \cite{Ali:2009zq, Ali:2011fa, Das:2010zf}. This quantization of area ($N=2$) and volume  ($N=3$) are given by 
\bea
V_N \equiv
\prod_{k=1}^N \frac{L_k}{a_0 \ell_{Pl}}
~.\hspace{1.5ex}
\eea
In order to have a relativistic discreteness of space, two authors of this paper studied the relativistic anisotropic spin-orbit interaction in \cite{Ali:2021oml}. In general, the spin-orbit interaction happens between a particle's spin and force potential. As we discussed in the introduction, we consider electromagnetic force as the dominating force. For that purpose, we write Dirac's equation in an electromagnetic field which has a form that respect Lorentz symmetry and gauge symmetry. It is given by the following equation

\bea
H \psi &=&  \le(c\, \vec \a \cdot \vec p -
e   \vec \a \cdot   \vec A+ e \phi + \b mc^2 \ri) \psi (\vec r)\nn \\
&=& E\psi (\vec r)
\la{ham2}
\eea
where $\vec A$ is a vector potential of EM field and $\phi$ is its scalar potential. In this paper, we seek a solution of a possible equivalency between the GUP corrected term (that implies discreteness) and electromagnetic interaction (that is Lorentz and gauge invariant). This happens if we set the correction implied GUP  in Eq. (\ref{ham1}) equal to the correction implied by EM field in Eq.(\ref{ham2}). This will give the following equation
\begin{equation}
    c a (\vec\a \cdot \vec p)^2 \psi= \le(e   \vec \a \cdot   \vec A+ e \phi\ri)  \psi
    \label{cojecture}
\end{equation}

For simplicity, we consider the one-dimensional case.

\begin{equation}
    \alpha_x^2 p_x^2~~\psi= \frac{e}{a c} (\alpha_x A_x +\phi) \psi
\end{equation}
where $\alpha_x =\big(\begin{smallmatrix}
  0 & \sigma_x\\
  \sigma_x & 0
\end{smallmatrix}\big)$ and $\sigma_x =\big(\begin{smallmatrix}
  0 & 1\\
  1 & 0
\end{smallmatrix}\big)$ is one of Pauli matrices. Since $\alpha_x^2=I$, we get
\begin{equation}
    -\hbar^2 \frac{\partial^2}{\partial x^2} \psi~ I= \frac{e}{ac}\le( \alpha_x A_x+ \phi \ri) \psi \label{GUPEM}
\end{equation}
If we consider the diagonal part of Eq. (\ref{GUPEM}), we get the following equation
\begin{equation}
    \hbar^2 \frac{\partial^2}{\partial x^2} \psi(x)=- \frac{e}{ac} \phi(x) \psi(x) \label{GUPEM1}
\end{equation}
The off-diagonal part will give an equation in terms of the EM potential vector $A_x$, which we will study later in a different project. Here we only work with the scalar potential $\phi$ which appears in the diagonal part. Consider the Coulomb potential case for $\phi(x)$ that is given by 
\begin{equation}
    \phi_x=\frac{e}{4 \pi \epsilon_0 x} \label{coloumb}
\end{equation}
where $\epsilon_0$ is the permittivity of free space. If we substitute Eq. (\ref{coloumb}) in Eq. (\ref{GUPEM1}), we get the following equation

\begin{equation}
 \frac{\partial^2}{\partial x^2} \psi(x)+ \frac{\beta}{\hbar a} \frac{\psi(x)}{x}=0 \label{GUPEM2}
\end{equation}
where $\beta=\frac{e^2}{4 \pi \epsilon_0 \hbar c}$ is the fine-structure constant. The solution of Eq. (\ref{GUPEM2}) is given by

\begin{equation}
    \psi \left( x \right) = \left( {{\sl Y}_{1}\left(2\,\sqrt {\frac{\beta }{\hbar a}}\sqrt {x}
\right)}{\it \_C_1}+{{\sl J}_{1}\left(2\,\sqrt {\frac{\beta}{\hbar a}}\sqrt {x}\right)}{
\it \_C_2} \right) \sqrt {x} \label{solution11}
\end{equation}
where ${\sl Y}_{1}$ and  ${\sl J}_{1}$ are Bessel functions, whereas $C_{1}$ and $C_{2}$ are arbitrary constants. This solution can be plotted as follows.
\begin{figure}[ht]
\begin{center}
\includegraphics[width=0.7\textwidth]{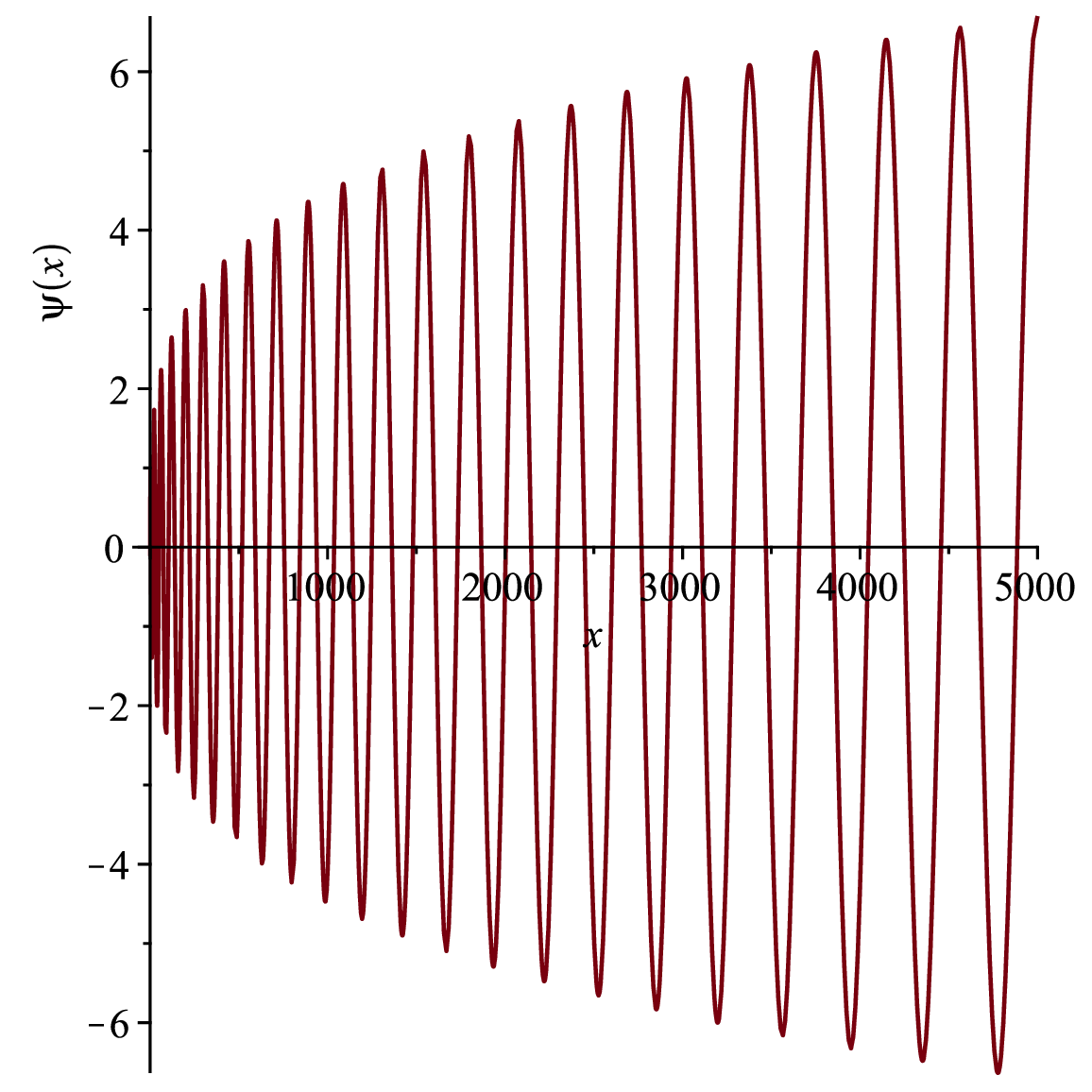}
\end{center}
\vspace{-0.5cm}
\caption{Plot for the wave function solution in Eq. (\ref{solution11}) for the one dimensional case.}
\label{solutionfig}
\end{figure}
The wave function solution given in Eq. (\ref{solution11}) is presented in Fig.(\ref{solutionfig}) which fully describes a discrete picture of the space that respects Lorentz symmetry and gauge symmetry. We know that the state of a quantum system is completely described by its wave function. The appearance of Bessel functions in the problems of wave propagation is quite common and expected. It is well-known that the wave function of a quantum system can be interpreted as a probability amplitude. So in the plot $\psi(x)$ may be interpreted as the probability of finding the particle at the point $x$. From Fig.(\ref{solutionfig}) we see that the wave grows in amplitude with $x$. This means that the probability of finding the particle increases with $x$, i.e. as we move away from the origin. As we move towards the origin along the $x-$axis the wave function gradually collapses showing that the chance of finding the particle decreases as we approach the origin. The above conceptualization naturally gives a discrete picture of space with the particle being delocalized in the wave and the curve showing its probable location. This solution may explain the crystal and quasicrystal structures that are observed in nature \cite{kortan1990real, holland1993observation, vacaru2018space}. In the next section, we generalize our computations in two and three dimensions. 

\section{Dirac equation in two dimensions}
Considering a two dimensional scenario the eqn.(\ref{cojecture}) leads us to,
\begin{equation}\label{2dirac}
\left(\alpha_{x}p_{x}+\alpha_{y}p_{y}\right)^{2}\psi(x,y)=\frac{e}{ac}\left(\alpha_{x}A_{x}+\alpha_{y}A_{y}+\phi\right)\psi(x,y)
\end{equation}
where $\alpha_y =\big(\begin{smallmatrix}
  0 & \sigma_y\\
  \sigma_y & 0
\end{smallmatrix}\big)$ and $\sigma_y =\big(\begin{smallmatrix}
  0 & -i\\
  i & 0
\end{smallmatrix}\big)$ is a Pauli matrix. $\alpha_x$ and $\sigma_{x}$ have already been defined before. Using the relations $\alpha_{x}^{2}=\alpha_{y}^{2}=1$ and the anti-commutative relation $\alpha_{x}\alpha_{y}+\alpha_{y}\alpha_{x}=0$ the above equation may be put in the form,
\begin{equation}\label{2d1}
-\hbar^{2}\left(\frac{\partial^2}{\partial x^2}+\frac{\partial^2}{\partial y^2} \right)\psi(x,y)=\frac{e}{ac}\left(\alpha_{x}A_{x}+\alpha_{y}A_{y}+\phi\right)\psi(x,y)
\end{equation}
Considering the diagonal part of eqn.(\ref{2d1}) we get,
\begin{equation}\label{2d2}
-\hbar^{2}\left(\frac{\partial^2}{\partial x^2}+\frac{\partial^2}{\partial y^2} \right)\psi(x,y)=\frac{e\phi}{ac}\psi(x,y)
\end{equation}
Here we consider the two-dimensional Coulomb potential for $\phi$ given by
\begin{equation}\label{col1}
\phi(x,y)=\frac{1}{4\pi\epsilon_{0}}\frac{e}{\sqrt{x^{2}+y^{2}}}
\end{equation}
It should be noted here that the Coulomb potential has different forms for different dimensions and different conditions and so the above consideration might not be totally exact. But here we are considering an isotropic charge distribution. Motivated by this symmetry we are considering simplified analogical forms of the Coulomb potential for all the dimensions. This will help us to find analytical solutions to an otherwise extremely complicated mathematical system. Using the above value of Coulomb potential in eqn.(\ref{2d2}) we get,
\begin{equation}\label{mixed}
\left(\frac{\partial^2}{\partial x^2}+\frac{\partial^2}{\partial y^2} \right)\psi(x,y)=-\frac{e}{ac\hbar^{2}}\left(\frac{e}{4\pi\epsilon_{0}\sqrt{x^{2}+y^{2}}}\right)\psi(x,y)
\end{equation}
Unfortunately, there is no general solution for the above equation. So we have to consider special cases for the coulomb potential to derive meaningful solutions from the equation. We consider the following two cases.\\

\subsection{Case I}

Here we consider one dimensional Coulomb potential only in $x$ given by $\phi(x,y)=\phi_x=\frac{e}{4 \pi \epsilon_0 x}$. Using this in eqn.(\ref{mixed}) we get,
\begin{equation}\label{onlyx}
\left(\frac{\partial^2}{\partial x^2}+\frac{\partial^2}{\partial y^2} \right)\psi(x,y)=-\frac{e}{ac\hbar^{2}}\left(\frac{e}{4\pi\epsilon_{0}x}\right)\psi(x,y)
\end{equation}
Now we apply the method of variable separation to solve the above-simplified equation.  For this, we consider the following variable separated form of the two-dimensional wave function $\psi(x,y)$,
\begin{equation}\label{sep}
\psi(x,y)=\psi_{1}(x)~\psi_{2}(y)
\end{equation}
Using the above form in eqn.(\ref{onlyx}) we get,
\begin{equation}
\frac{\psi_{1}''(x)}{\psi_{1}(x)}+\frac{\beta}{a\hbar}\frac{1}{x}=-\frac{\psi_{2}''(y)}{\psi_{2}(y)}=\lambda
\end{equation}
where prime denotes derivative with respect to the argument, $\lambda$ is an arbitrary constant, and $\beta$ is the fine-structure constant defined before. Now we can break the above equation into two parts for $x$ and $y$ and solve them separately. Separating the above equation we get,
\begin{equation}\label{part1}
\frac{\psi_{1}''(x)}{\psi_{1}(x)}+\frac{\beta}{a\hbar}\frac{1}{x}=\lambda
\end{equation}
and
\begin{equation}\label{part2}
\frac{\psi_{2}''(y)}{\psi_{2}(y)}=-\lambda
\end{equation}
Solving eqn.(\ref{part1}), we get

$$\psi_{1}(x)=\frac{xe^{-\sqrt{\lambda}~x}}{a\hbar}\left(C_{3}~U\left[1-\frac{\beta}{2a\hbar\sqrt{\lambda}},2,2\sqrt{\lambda}~x\right]\right.$$
\begin{equation}\label{soln1}
\left.+C_{4}~1F1\left[1-\frac{\beta}{2a\hbar\sqrt{\lambda}},2,2\sqrt{\lambda}~x\right]\right)
\end{equation}
where $U$ and $1F1$ are hypergeometric functions, whereas $C_{3}$ and $C_{4}$ are arbitrary constants.
Now solving eqn.(\ref{part2}) we get,
\begin{equation}\label{soln2}
\psi_{2}(y)=C_{5}\cos\left(\sqrt{\lambda}~y\right)+C_{6}\sin\left(\sqrt{\lambda}~y\right)
\end{equation}
where $C_{5}$ and $C_{6}$ are arbitrary constants. Now using eqns.(\ref{soln1}) and (\ref{soln2}) in eqn.(\ref{sep}) we get the solution for the wave function as,
$$\psi(x,y)=\frac{xe^{-\sqrt{\lambda}~x}}{a\hbar}\left(C_{3}~U\left[1-\frac{\beta}{2a\hbar\sqrt{\lambda}},2,2\sqrt{\lambda}~x\right]\right.$$

$$\left.+C_{4}~1F1\left[1-\frac{\beta}{2a\hbar\sqrt{\lambda}},2,2\sqrt{\lambda}~x\right]\right)\left[C_{5}\cos\left(\sqrt{\lambda}~y\right)\right.$$
\begin{equation}\label{part1part2}
\left.+C_{6}\sin\left(\sqrt{\lambda}~y\right)\right]
\end{equation}

\begin{figure}[ht]
\begin{center}
\includegraphics[width=0.7\textwidth]{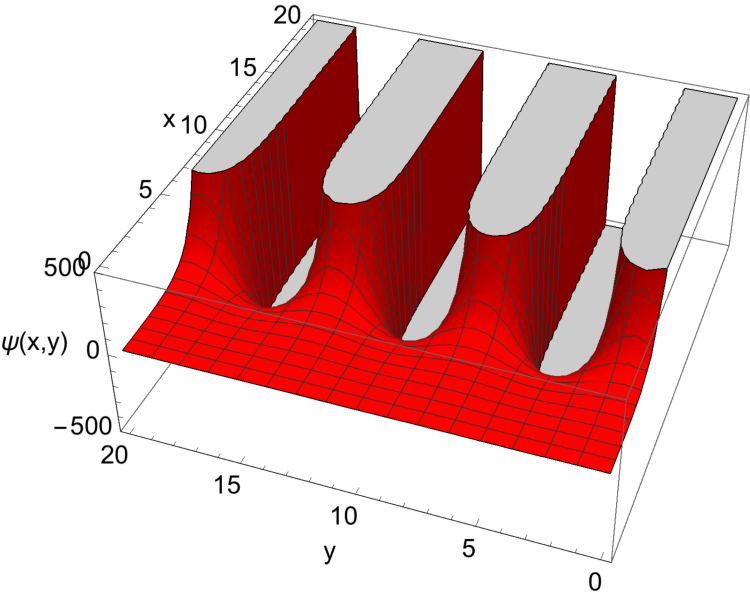}
\end{center}
\vspace{-0.5cm}
\caption{Plot for wave function solution of Eq. (\ref{part1part2}) for the two dimensional case. This is a graphical representation and the constants are chosen to be unity. Plot not to scale.}
\label{solutionfig2D}
\end{figure}

The above solution has been plotted in Fig.(\ref{solutionfig2D}). Just like Fig.(\ref{solutionfig}), interpreting the wave function as the probability amplitude of the two-dimensional quantum system, we get an idea of the position of the particle in the wave. The peaks represent the positions where there is a high probability of finding the quantum particle. At other places, there is a comparatively lower chance of finding the particle. This gives a clear and complete picture of the discrete nature of the two-dimensional system.

\subsection{Case II}
Here we consider one dimensional Coulomb potential only in $y$ given by $\phi(x,y)=\phi_y=\frac{e}{4 \pi \epsilon_0 y}$. However as far as the solution is concerned this case will be similar to the previous case with just $x$ and $y$ being interchanged among themselves. So the solution for this case may be directly written from eqn.(\ref{part1part2}) as below,
$$\psi(x,y)=\frac{ye^{-\sqrt{\lambda}~y}}{a\hbar}\left(C_{3}~U\left[1-\frac{\beta}{2a\hbar\sqrt{\lambda}},2,2\sqrt{\lambda}~y\right]\right.$$

$$\left.+C_{4}~1F1\left[1-\frac{\beta}{2a\hbar\sqrt{\lambda}},2,2\sqrt{\lambda}~y\right]\right)\left[C_{5}\cos\left(\sqrt{\lambda}~x\right)\right.$$
\begin{equation}\label{part1part22}
\left.+C_{6}\sin\left(\sqrt{\lambda}~x\right)\right]
\end{equation}
The solution preserves the qualitative features of the solution in case I, the only difference being the interchange of directional properties. The discrete crystal structure will be similar in this case as found in case I, with the directional properties between $x$ and $y$ interchanged. 

\section{Dirac equation in three dimensions}
Now the above formalism can be easily extended to the three-dimensional Dirac equation. In three dimensions eqn.(\ref{2dirac}) can be extended to the form,

$$\left(\alpha_{x}p_{x}+\alpha_{y}p_{y}+\alpha_{z}p_{z}\right)^{2}\psi(x,y,z)=\frac{e}{ac}\left(\alpha_{x}A_{x}+\alpha_{y}A_{y}\right.$$
\begin{equation}\label{3dirac}
\left.+\alpha_{z}A_{z}+\phi\right)\psi(x,y,z)
\end{equation}
where $\alpha_z =\big(\begin{smallmatrix}
  0 & \sigma_z\\
  \sigma_z & 0
\end{smallmatrix}\big)$ and $\sigma_z =\big(\begin{smallmatrix}
  1 & 0\\
  0 & -1
\end{smallmatrix}\big)$ is a Pauli matrix. The other matrices have been defined before. Using the relations $\alpha_{x}^{2}=\alpha_{y}^{2}=\alpha_{z}^{2}=1$ and the anti-commutative relations $\alpha_{x}\alpha_{y}+\alpha_{y}\alpha_{x}=\alpha_{y}\alpha_{z}+\alpha_{z}\alpha_{y}=\alpha_{z}\alpha_{x}+\alpha_{x}\alpha_{z}=0$ the above equation may be put in the form,

$$-\hbar^{2}\left(\frac{\partial^2}{\partial x^2}+\frac{\partial^2}{\partial y^2}+\frac{\partial^2}{\partial z^2} \right)\psi(x,y,z)=\frac{e}{ac}\left(\alpha_{x}A_{x}+\alpha_{y}A_{y}\right.$$
\begin{equation}\label{3d1}
\left.+\alpha_{z}A_{z}+\phi\right)\psi(x,y,z)
\end{equation}
Considering the diagonal part of eqn.(\ref{3d1}) we get,
\begin{equation}\label{3d2}
-\hbar^{2}\left(\frac{\partial^2}{\partial x^2}+\frac{\partial^2}{\partial y^2}+\frac{\partial^2}{\partial z^2} \right)\psi(x,y,z)=\frac{e\phi}{ac}\psi(x,y,z)
\end{equation}
Since it is known from the analysis in two dimensions that a general solution cannot be obtained by considering a three-dimensional analog of the Coulomb potential given in eqn.(\ref{col1}), we directly move on to consider special cases with a one-dimensional form of potentials.

\subsection{Case I}
Here we consider one dimensional Coulomb potential only in $x$ given by $\phi(x,y,z)=\phi_x=\frac{e}{4 \pi \epsilon_0 x}$. Using this in eqn.(\ref{3d2}) we get,
\begin{equation}\label{onlyx2}
\left(\frac{\partial^2}{\partial x^2}+\frac{\partial^2}{\partial y^2}+\frac{\partial^2}{\partial z^2} \right)\psi(x,y,z)=-\frac{e}{ac\hbar^{2}}\left(\frac{e}{4\pi\epsilon_{0}x}\right)\psi(x,y,z)
\end{equation}
Now we apply the method of variable separation to solve the above-simplified equation.  For this, we consider the following variable separated form of the three-dimensional wave function $\psi(x,y,z)$,
\begin{equation}\label{sep2}
\psi(x,y,z)=\psi_{1}(x)~\psi_{2}(y)~\psi_{3}(z)
\end{equation}
Using this in eqn.(\ref{onlyx2}) we get three variable separated equations as,
\begin{equation}\label{3eq1}
\frac{\psi_{1}''(x)}{\psi_{1}(x)}+\frac{\beta}{a\hbar}\frac{1}{x}=\gamma
\end{equation}

\begin{equation}\label{3eq2}
\frac{\psi_{2}''(y)}{\psi_{2}(y)}=\mu-\gamma
\end{equation}
and
\begin{equation}\label{3eq3}
\frac{\psi_{3}''(z)}{\psi_{3}(z)}=-\mu
\end{equation}
where $\gamma$ and $\mu$ are arbitrary constants.
Now the solution of the eqn.(\ref{3eq1}) can be written analogically using eqn.(\ref{soln1}) as,
$$\psi_{1}(x)=\frac{xe^{-\sqrt{\gamma}~x}}{a\hbar}\left(C_{7}~U\left[1-\frac{\beta}{2a\hbar\sqrt{\gamma}},2,2\sqrt{\gamma}~x\right]\right.$$
\begin{equation}\label{soln11}
\left.+C_{8}~1F1\left[1-\frac{\beta}{2a\hbar\sqrt{\gamma}},2,2\sqrt{\gamma}~x\right]\right)
\end{equation}
where $C_{7}$ and $C_{8}$ are arbitrary constants. Similarly the solution of eqns.(\ref{3eq2}) and (\ref{3eq3}) can be written analogically from eqn.(\ref{soln2}) respectively as,
\begin{equation}\label{soln22}
\psi_{2}(y)=C_{9}\cos\left(\sqrt{\gamma-\mu}~y\right)+C_{10}\sin\left(\sqrt{\gamma-\mu}~y\right)
\end{equation}
and
\begin{equation}\label{soln33}
\psi_{3}(z)=C_{11}\cos\left(\sqrt{\mu}~z\right)+C_{12}\sin\left(\sqrt{\mu}~z\right)
\end{equation}
where $C_{9}$, $C_{10}$, $C_{11}$ and $C_{12}$ are arbitrary constants. Finally using eqns.(\ref{soln11}), (\ref{soln22}) and (\ref{soln33}) in eqn.(\ref{sep2}) we get the solution of the wave equation as,

$$\psi(x,y,z)=\frac{xe^{-\sqrt{\gamma}~x}}{a\hbar}\left(C_{7}~U\left[1-\frac{\beta}{2a\hbar\sqrt{\gamma}},2,2\sqrt{\gamma}~x\right]\right.$$

$$\left.+C_{8}~1F1\left[1-\frac{\beta}{2a\hbar\sqrt{\gamma}},2,2\sqrt{\gamma}~x\right]\right)\left[C_{9}\cos\left(\sqrt{\gamma-\mu}~y\right)\right.$$

\begin{equation}\label{3dsolnf}
\left.+C_{10}\sin\left(\sqrt{\gamma-\mu}~y\right)\right]\left[C_{11}\cos\left(\sqrt{\mu}~z\right)+C_{12}\sin\left(\sqrt{\mu}~z\right)\right]
\end{equation}
This gives the final solution of the wave function in a variable-separated form when the Coulomb potential is considered to be present only in the x-direction. This solution gives the discrete nature of the space in three dimensions. The role played by the hypergeometric functions in generating this discrete crystal-like spatial structure is crucial and can be considered as the central idea in this work.

\subsection{Case II}
Here we consider one dimensional Coulomb potential only in $y$ given by $\phi(x,y,z)=\phi_y=\frac{e}{4 \pi \epsilon_0 y}$. As we have already seen in the case of two dimensions, these cases can be written analogically using the solution of case I due to their symmetric nature. So in this case we can get the solution using eqn.(\ref{3dsolnf}) by interchanging $x$ and $y$ among themselves as below,

$$\psi(x,y,z)=\frac{ye^{-\sqrt{\gamma}~y}}{a\hbar}\left(C_{7}~U\left[1-\frac{\beta}{2a\hbar\sqrt{\gamma}},2,2\sqrt{\gamma}~y\right]\right.$$

$$\left.+C_{8}~1F1\left[1-\frac{\beta}{2a\hbar\sqrt{\gamma}},2,2\sqrt{\gamma}~y\right]\right)\left[C_{9}\cos\left(\sqrt{\gamma-\mu}~x\right)\right.$$

\begin{equation}\label{3dsolnff}
\left.+C_{10}\sin\left(\sqrt{\gamma-\mu}~x\right)\right]\left[C_{11}\cos\left(\sqrt{\mu}~z\right)+C_{12}\sin\left(\sqrt{\mu}~z\right)\right]
\end{equation}
This form of this solution and its qualitative features are quite similar to those of the solution given in the previous case, with the features of $x$ and $y$ directions interchanged. Thus overall the solution will be different even though they look quite similar. Physically the crystalline structures will also be different in the two cases since the components in various directions are different.

\subsection{Case III}
Here we consider one dimensional Coulomb potential only in $z$ given by $\phi(x,y,z)=\phi_z=\frac{e}{4 \pi \epsilon_0 z}$. Just like the previous case, here the solution may be written using eqn.(\ref{3dsolnf}) by simply interchanging the variables $x$ and $z$ among themselves as given below,

$$\psi(x,y,z)=\frac{ze^{-\sqrt{\gamma}~z}}{a\hbar}\left(C_{7}~U\left[1-\frac{\beta}{2a\hbar\sqrt{\gamma}},2,2\sqrt{\gamma}~z\right]\right.$$

$$\left.+C_{8}~1F1\left[1-\frac{\beta}{2a\hbar\sqrt{\gamma}},2,2\sqrt{\gamma}~z\right]\right)\left[C_{9}\cos\left(\sqrt{\gamma-\mu}~y\right)\right.$$

\begin{equation}\label{3dsolnfff}
\left.+C_{10}\sin\left(\sqrt{\gamma-\mu}~y\right)\right]\left[C_{11}\cos\left(\sqrt{\mu}~x\right)+C_{12}\sin\left(\sqrt{\mu}~x\right)\right]
\end{equation}
The qualitative features of the wave functions in the three-dimensional case will be almost similar to the wave function in the two-dimensional case as is evident from Fig.(2). The space will be discretized, but in a higher dimension. The role of electric charge in generating this discrete nature of space is crucial and clearly evident in this work.

\section{Discussion \& Conclusion}
In this work, we have derived a discrete picture of space that respects the Lorentz symmetry and the gauge symmetry. An equivalence between the linear GUP correction term and the electromagnetic interaction term is established in the Dirac equation. A wavefunction is obtained as a solution to the Dirac equation, which gave a relativistic quantum picture of space driven by the electromagnetic interaction. The solution of the wave function is investigated in one, two, and three dimensions. For the case of one dimension, the system admits a general solution for the generated partial differential equation. This solution is given in terms of Bessel functions, which is a special function originating from the solution of various partial differential equations. The solutions of differential equations as special functions play a very important role in quantum physics. These special functions (mainly Bessel function) are eigenfunctions of Hermitian operators, which are of special importance in quantum problems in calculating the real expectation values of a dynamical variable. Moreover, Bessel functions are also important for different problems of wave propagation and static potentials. We have plotted the obtained wavefunction in Fig.(1). From the figure the discrete nature of the space is clearly evident.

Next, we have explored the solutions for the two and three-dimensional cases. Unfortunately that for both these cases there are no general solutions for the obtained partial differential equations. So we have resorted to special cases and sought solutions for such scenarios. In these particular scenarios, we have considered one-dimensional Coulomb potential respectively in the $x$ and $y$ directions for two dimensions and $x$, $y$, and $z$ directions for the three dimensions. It is seen that for both cases we have obtained solutions in terms of hypergeometric functions, which are special functions represented by the hypergeometric series. Just like Bessel functions, hypergeometric functions also originate as solutions of differential equations (second-order linear ordinary differential equations). It is to be noted here that the solutions for the various subcases (Coulombic potential in $x$ or $y$ or $z$ direction) are symmetric in these variables $x$, $y$, and $z$. So the qualitative features of a particular case are preserved for all the other cases, the difference being the appearance of the features in a different direction. So we have plotted the wavefunction obtained in the first case of two dimensions in Fig.(2). It is seen that it shows a perfect scenario of discretization of space. The plots for the other cases will be similar since the qualitative features of the other solutions are similar, and hence they are skipped. The discreteness presented in this work may explain the crystal and the quasicrystal structures observed in nature at different energy scales.

\section{Acknowledgement}
P.R. acknowledges the Inter-University Centre for Astronomy and Astrophysics (IUCAA), Pune, India for granting visiting
associateship.

\bibliographystyle{utcaps}
\bibliography{ref.bib}{}

\end{document}